\documentclass[prd,a4paper]{revtex4}
\usepackage{epsfig}

\newcommand{\be}{\begin{equation}}
\newcommand{\ee}{\end{equation}}
\newcommand{\bea}{\begin{eqnarray}}
\newcommand{\eea}{\end{eqnarray}}
\newcommand{\bean}{\begin{eqnarray*}}
\newcommand{\eean}{\end{eqnarray*}}

\newcommand{\gapproxeq}{\lower
.7ex\hbox{$\;\stackrel{\textstyle >}{\sim}\;$}}
\newcommand{\lapproxeq}{\lower
.7ex\hbox{$\;\stackrel{\textstyle <}{\sim}\;$}}

\begin{document}

\bibliographystyle{unsrt}

\title{\bf A coherent study of $\chi_{c0,2}\to VV$, $PP$ and $SS$ }

\author{Qiang Zhao$^{1,2}$\footnote{e-mail: qiang.zhao@surrey.ac.uk} 
}

\affiliation{1) Institute of High Energy Physics,
Chinese Academy of Sciences, Beijing, 100049, P.R. China}

\affiliation{2)Department of Physics,
University of Surrey, Guildford, GU2 7XH, United Kingdom}

\date{\today}

\begin{abstract}

We investigate the decays of $\chi_{c0,2}$ into vector meson pairs ($VV$), 
pseudoscalar pairs ($PP$), and scalar pairs ($SS$) in a general factorization 
scheme. 
The purpose is to clarify the role played by 
the OZI-rule violations and  SU(3) flavour breakings 
in the decay transitions, and their correlations with the final state meson 
wavefunctions. For $\chi_{c0,2}\to VV$ and $PP$, we obtain 
an overall self-contained description of the experimental data. 
Applying this factorization to 
$\chi_{c0}\to f_0^i f_0^j$, where $i, \ j=1, \ 2, \ 3$ denotes 
$f_0(1710)$, $f_0(1500)$ and $f_0(1370)$, respectively, we find that 
specific patterns will arise from the model predictions for 
the decay branching ratio magnitudes, and useful information about the 
structure of those three scalars can be abstracted.

\end{abstract}

\maketitle


Charmonium decays provide a great opportunity for studying the gluon dynamics 
at relatively low energies. In line with the lattice QCD predictions for the 
existence of the lowest glueball state ($0^{++}$) at 1.45$\sim$ 1.75 GeV~\cite{ukqcd,mp}, 
one of the most important motivations to study charmonium decays 
is to search for glueball states and fill in the missing link of QCD.  
During the past few years, 
enriched information from experiments reveals a rather crowded scalar meson 
spectrum at 1$\sim$ 2 GeV. For example, 
three $f_0$ states, $f_0^{1,2,3}$ ($=f_0(1710), \ f_0(1500), \ f_0(1370)$), 
are observed in both $pp$ scattering and $p\bar{p}$ annihilations with 
different decay modes into pseudoscalar pairs~\cite{wa102,anisovich,amsler,bugg}. 
They are also confirmed by the BES data 
in $J/\psi\to \omega f_0^i$ and $\phi f_0^i$~\cite{bes-phi,bes-plb,bes-f0(1500)}. 
Moreover, another scalar $f_0(1790)$, 
which is distinguished from $f_0(1710)$, is also seen 
at BES~\cite{jin}. In contrast with the $f_0(1710)$, 
$f_0(1790)$ does not obviously couple to $K\bar{K}$, while its coupling 
to $\pi\pi$ is quite strong. 
Since all these states cannot be simply accommodated into a $Q\bar{Q}$ 
meson configuration, observation of such a rich scalar-meson spectrum could be 
signals for exotic states beyond simple $Q\bar{Q}$ 
configurations~\cite{close-tornq,close-ichep04}. 
The fast-growing database also 
allows us to develop QCD phenomenologies from which 
we can gain insights into the glueball 
production mechanism~\cite{close-amsler,cfl,close-kirk,close-zhao-f0,close-zhao-glueball}. 

So far, it is still a mystery 
that the glueball states have remained hidden 
from experiments for such a long time~\cite{close-88,close-ichep04}. 
One possibility could be that a pure glueball state cannot survive a long 
enough lifetime before it transits into $Q\bar{Q}$ pairs. 
Because of this, a glueball would be more likely to mix with nearby $Q\bar{Q}$. 
As a result, it should be more sensible to look for a glueball-$Q\bar{Q}$ mixing 
state, instead of a pure glueball. 
Such a mixing scheme, on the one hand, 
indeed highlights some characters of the meson properties,  
and fits in the experimental data quite well~\cite{amsler,cfl,close-zhao-f0}.
It seems also  promising in accounting for the crowded scalar 
meson spectrum at 1$\sim$ 2 GeV~\cite{close-zhao-f0}. 
On the other hand, it raises questions 
about the role played by the OZI rule~\cite{ozi} 
in the glueball production mechanism, which 
needs understanding in a much broader context. 
In Ref.~\cite{zhao-zou}, it is shown that the idea of intermediate virtual 
meson exchanges~\cite{lipkin,isgur-geiger,lipkin-zou} 
can provide a dynamic explanation for the large OZI-rule violations 
in $J/\psi\to \omega f_0(1710)$ and $\phi f_0(1710)$. 
It implies that correlations of OZI-rule violations 
could be an important dynamic process in charmonium decays. 
Such correlations should not be exclusively restricted to the scalar meson productions 
in $J/\psi$ decays.  
It could also contribute to other processes~\cite{seiden}, 
such as $\chi_{c0,2}$ hadronic decays. 
Therefore, a systematic investigation of $\chi_{c0,2}\to VV$, $PP$ and $SS$ 
is necessarily useful for clarifying the role played by the OZI-rule violations.

The other relevant issue in the charmonium decays into light hadrons 
is the SU(3) flavour symmetry breakings. In many cases, the flavour-blind 
assumption turns out to be a good approximation for the gluon-$Q\bar{Q}$ 
couplings. However, due to the presence of OZI-violations, an explicit 
consideration of the SU(3) flavour symmetry breaking will be useful for 
disentangling correlations from different mechanisms. 

In this work, we will present a factorization scheme for $\chi_{c0,2}\to VV$, 
$PP$, and $SS$. Our purpose is to disentangle the roles played the 
OZI-rule violations and SU(3) flavour symmetry breakings, which will correlate 
with the final state meson wavefunctions in $\chi_{c0,2}$ decays.  
By this study, we hope to abstract additional information about the structure of 
the scalar mesons $f_0^i$. It can be reflected by 
specific patterns arising from the decay branching ratio magnitudes due to 
such correlations.

\begin{figure}
\begin{center}
\epsfig{file=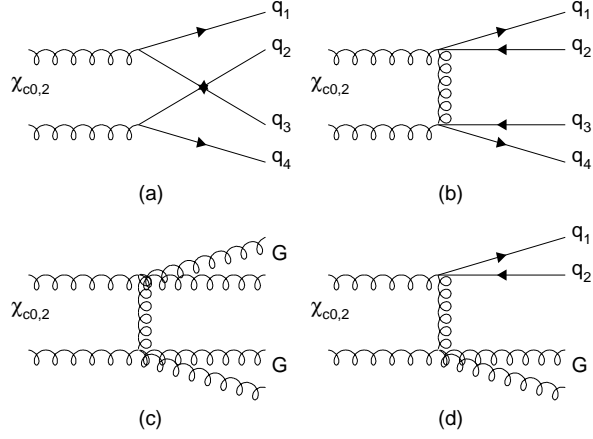, width=10cm,height=8.cm}
\caption{Schematic pictures for the decays of $\chi_{c0,2}$ into 
meson pairs via the production of different components. 
}
\protect\label{fig-1}
\end{center}
\end{figure}

To proceed, we first distinguish the singly OZI disconnected (SOZI)
processes and doubly OZI disconnected (DOZI) ones by considering transitions 
illustrated by Fig.~\ref{fig-1}. 
The SOZI transition [Fig.~\ref{fig-1}(a)] hence can be expressed as 
\be
\label{sozi}
\langle q_1\bar{q}_2 q_3\bar{q}_4 | V_0 | gg\rangle = 
g_{\langle 13\rangle} g_{\langle 24\rangle} \ ,
\ee
where $V_0$ is the interaction potential, 
and $|gg\rangle$ denotes the minimum two-gluon 
radiations in $c\bar{c}$ annihilation; 
$g_{13}$ and $g_{24}$ are the coupling strengths.   
For the non-strange coupling, we denote $g_{13}=g_{24}=g_0$.
To include the SU(3) flavour symmetry breaking effects, 
we introduce 
$R\equiv \langle s\bar{s}|V_0|g\rangle /\langle u\bar{u}|V_0|g\rangle
=\langle s\bar{s}|V_0|g\rangle /\langle d\bar{d}|V_0|g\rangle$, 
for which $R=1$ refers to the SU(3) flavour symmetry limit. 

The DOZI process of Fig.~\ref{fig-1}(b) describes the transition that 
the two $q\bar{q}$ pairs are created and recoil without quark exchanges. 
The amplitude can be expressed as
\be
\label{dozi-1}
\langle q_1\bar{q}_2 q_3\bar{q}_4 | V_1 | gg\rangle = 
g^\prime_{\langle 12\rangle}g^\prime_{\langle 34\rangle} 
= r g_{\langle 12\rangle}g_{\langle 34\rangle} \ ,
\ee
where $V_1$ denotes the interaction potential, and  
parameter $r$ is defined as the relative 
strength between the SOZI and DOZI transition amplitudes. 
It is introduced to take into account the DOZI effects.

The two gluons from the $c\bar{c}$ annihilation can also couple 
to the glueball components via Fig.~\ref{fig-1}(c). 
The transition amplitude can be expressed as 
\be
\label{dozi-2}
\langle GG| V_2| gg\rangle =g^{\prime\prime}g^{\prime\prime} = t g_0^2  \ ,
\ee
where $V_2$ is the interaction potential and $g^{\prime\prime}$ 
is the gluon coupling to the scalar glueball; Parameter $t$ is introduced 
to account for its coupling strength relative 
to the SOZI process. 

Another possible process in the decay of $\chi_{c0,2}$ 
is via Fig.~\ref{fig-1}(d), where the two gluons will couple 
to $q\bar{q}$ and glueball, respectively. 
According to Eqs.~\ref{dozi-1} and \ref{dozi-2}, 
its relative coupling strength to the SOZI process 
can be written as
\be
\langle q_1\bar{q}_2 G| V_3| gg\rangle 
= \sqrt{r t} g_0^2\ .
\ee

With the above factorization scheme, we consider 
a transition of $\chi_{c0,2}\to M^iM^j$, 
where $M^i$ and $M^j$ denote 
two final-state quarkonia such as $VV$, $PP$, or $f_0^if_0^j$. 
Firstly, in a general basis we assume a wavefunction for $I=0$ $M^i$:
\be
M^i= x_i |G\rangle + y_i |s\bar{s}\rangle +z_i| n\bar{n}\rangle \ ,
\ee
where $|G\rangle$, $|s\bar{s}\rangle$, and 
$ | n\bar{n}\rangle =|u\bar{u}+d\bar{d}\rangle/\sqrt{2}$
are the pure glueball, and flavor singlet $q\bar{q}$ components, respectively;  
Coefficients 
$x_i$, $y_i$ and $z_i$ denote the mixing angles of those components, 
and satisfy the unitarity condition. 

The transition amplitude for $\chi_{c0,2}\to M^iM^j$ can be written as
\bea
&& \langle M^iM^j| (V_0+ V_1 +V_2 +V_3)| gg\rangle \nonumber\\
&=& \langle ( x_i G + y_i s\bar{s} + z_i n\bar{n}) 
( x_j G + y_j s\bar{s} + z_j n\bar{n}) | (V_0+ V_1 +V_2 +V_3)| gg\rangle \ .
\eea
Notice that the potential $V_0$ will only allow the transitions 
to $I=0$ meson pairs, i.e. transitions of 
$\langle n\bar{n}s\bar{s}| V_1| gg\rangle 
=\langle s\bar{s}n\bar{n}| V_1| gg\rangle =0$. 
We can then decompose the above transition into
\bea
\label{trans-1}
&& \langle M^iM^j| (V_0+ V_1 +V_2 +V_3)| gg\rangle \nonumber\\
&=& g_0^2 [ x_i (t x_j +R \sqrt{rt} y_j +\sqrt{2rt} z_j)
+R y_i(\sqrt{rt} x_j + (1+r)R y_j +\sqrt{2} r z_j) \nonumber\\
&&+z_i(\sqrt{2rt} x_j + \sqrt{2} r R y_j +(1+2r) z_j)] \ .
\eea

For $\chi_{c0,2}$ decays into other octet meson pairs with $I\ne 0$, 
e.g. $K\bar{K}$, $\pi\pi$, etc,
we can see that only $V_0$ transition is allowed, and  
all the others are forbidden. This selection rule immediately requires 
that decay channels with $I\ne 0$ will not suffer from the OZI-rule 
violations. As a consequence, the OZI violation effects and 
SU(3) flavour symmetry breaking can be separated out. 
The explicit reduction of Eq.~\ref{trans-1} for $\chi_{c0,2}\to VV$ and $PP$
will be given later.  

In the calculation of the partial decay widths, 
a commonly useful form factor is applied:
\be
{\cal F}^2({\bf p})=p^{2l}\exp(-{\bf p}^2/8\beta^2) \ ,
\ee
where ${\bf p}$ and $l$ are the three momentum and relative angular momentum 
of the final-state mesons, respectively, in the $\chi_{c0,2}$ rest frame. 
We adopt $\beta=0.5$ GeV, which is the same as 
in Refs.~\cite{close-amsler,close-kirk,close-zhao-f0}. 
Such a form factor will largely account for the size effects from 
the spatial wavefunctions of the initial and final state mesons. 
After taking into account this, we assume that parameter $g_0$ 
will be an overall factor for a given initial charmonium.

i) $\chi_{c0,2}\to VV$

For  $\chi_{c0,2}\to \phi\phi$, $\omega\omega$, and $\omega\phi$, 
the transition amplitude reduces to simple forms due to their 
ideally mixed flavor wavefunctions, 
i.e. 
$\phi=s\bar{s}$, and $\omega=n\bar{n}=(u\bar{u}+d\bar{d})/\sqrt{2}$. 
This means that for $\phi$ meson, $x_i=z_i=0$ and $y_i=1$, and for 
$\omega$ meson, $x_j=y_j=0$ and $z_i=1$. Therefore, we have
\bea
\label{isospin-1}
\langle \phi\phi | \hat{V}| gg\rangle 
&=& g_0^2 R^2 (1+r)   \nonumber\\
\langle \omega\omega | \hat{V}| gg\rangle 
&=& g_0^2(1+2r)  \nonumber\\
\langle \omega\phi | \hat{V}| gg\rangle 
&=& g_0^2 r R \sqrt{2} \ ,
\eea
where we compactly write $V_0+ V_1+V_2+V_3$ as $\hat{V}$.

For  $\chi_{c0,2}\to K^*\bar{K^*}$ and $\rho\rho$, we have 
\bea
\label{isospin-2}
\langle K^{*+}K^{*-}|  \hat{V}| gg\rangle 
&=& g_0^2 R^2 \nonumber\\
\langle \rho^+\rho^- | \hat{V}| gg\rangle 
&=& g_0^2  \ . 
\eea
It can be shown that the amplitudes for other charge combinations 
are the same.

Interesting correlations arise from those decay amplitudes. 
Equation~\ref{isospin-2} shows that the decay of $\chi_{c0,2}\to\rho\rho$ 
is free of interferences from the DOZI processes. Nevertheless, 
it does not suffer from the possible SU(3) flavour symmetry breaking. 
Ideally, a measurement of this branching ratio will be useful for us to 
determine $g_0$. Unfortunately, experimental data for this channel are not available. 
The decay of $\chi_{c0,2}\to K^*\bar{K^*}$ also merits great advantages.  
It is also free of DOZI interferences, and the SU(3) breaking effects 
have simple correlation with the basic amplitude $g_0^2$. 
In contrast, Eq.~\ref{isospin-1} involves correlations from both 
OZI-rule violations and SU(3) breakings. 

So far, three channels have been measured for $\chi_{c0}$ decays
by BES collaboration~\cite{bes-99b,bes-04,bes-05}: 
$BR_{\chi_{c0}\to \phi\phi}=(1.0\pm 0.6)\times 10^{-3}$, 
$BR_{\chi_{c0}\to \omega\omega}
=(2.29\pm 0.58\pm 0.41)\times 10^{-3}$, and
$BR_{\chi_{c0}\to K^{*0}\bar{K^{*0}}}=(1.78\pm 0.34\pm 0.34)\times 10^{-3}$. 
This allows us to determine the parameters with very small $\chi^2$: 
\be
\label{para-value}
r=0.45\pm 0.48, \ \ \ R=0.90\pm 0.22, \ \ \ g_0=0.25\pm 0.06 \ \mbox{GeV}^{1/2}.
\ee
We note that the central values can be obtained by 
directly solving the equations for these three channels. 
Fitting the data can provide an estimate of the parameter uncertainties 
due to the experimental error bars. 
The results show that large errors of the data can lead to 
large uncertainties for the parameters, especially for $r$. 
To see more clearly their correlations, 
we take the fraction between 
the branching ratios for $\omega\omega$ and  $\phi\phi$ to derive:
\be
\label{para-vector}
r=\frac{R^2 C_0-1}{2-R^2 C_0} \ ,
\ee
where $C_0\equiv [p_\phi BR_{\chi_{c0}\to \omega\omega} 
{\cal F}^2({\bf p_\phi})
/p_\omega BR_{\chi_{c0}\to \phi\phi} 
{\cal F}^2({\bf p_\omega})]^{1/2}=1.6$. 
It shows that 
for a range of $1/C_0\le  R^2  < 2/C_0$, 
the above relation leads to $0\le r <+\infty$. 
In other words, the combined uncertainties 
from the SU(3) breaking and experimental 
uncertainties can result in rather significant changes to $r$. 
Therefore, improved measurements are still needed for the determination 
of $r$. 
Meanwhile, the value for $R$ suggests that the SU(3) breaking is about 20\%. 
This seems to be consistent with 
the success of SU(3) flavour symmetry in many circumstances. 

Adopting the central values for the parameters, 
we can predict the branching ratios for 
$\chi_{c0}\to \omega\phi$ and $\rho\rho$, 
and the results are listed in Table~\ref{tab-1}. 
We find a relatively smaller branching ratio for $\chi_{c0}\to \omega\phi$, 
i.e. $BR_{\chi_{c0}\to \omega\phi}=0.45\times 10^{-3}$.
This channel should be forbidden if the OZI-rule is respected. 
However, with the errors from Eq.~\ref{para-value}, the estimate 
of the root mean square error gives  
$1.5\times 10^{-3}$, which is the same order of magnitude as 
$\phi\phi$ and $\omega\omega$ channels. 
This implies that the OZI-rule violations could be significant. 
A direct measurement of $\omega\phi$ channel 
will put a better constraint on the OZI-rule violation effects.

The decay of $\chi_{c0}\to \rho\rho$ does not suffer the OZI rule, 
and its branching ratio is predicted to be sizeable. 
Notice that $\chi_{c0}$ has large branching ratios for 
$2(\pi^+\pi^-)$ and $\rho^0\pi^+\pi^-$~\cite{pdg2004}, the predicted value, 
$BR_{\chi_{c0}\to \rho\rho}=1.88\times 10^{-3}$, appears to be reasonable. 
A precise measurement of this channel is strongly desired for the determination 
of parameter $g_0$. Note that $g_0$ appears in the branching ratio 
with a power of four. The error for $g_0$ as in Eq.~\ref{para-value} 
can still lead to large uncertainties of $92\%$ for $BR_{\chi_{c0}\to \rho\rho}$.

For $\chi_{c2}\to VV$, so far, branching ratios for three channels 
have been available from BES~\cite{bes-99b,bes-04,bes-05}:
$BR_{\chi_{c2}\to \phi\phi}=(2.00\pm 0.55\pm 0.61)\times 10^{-3}$, 
$BR_{\chi_{c2}\to \omega\omega}
=(1.77\pm 0.47\pm 0.36)\times 10^{-3}$, and
$BR_{\chi_{c2}\to K^*\bar{K^*}}=(4.86\pm 0.56\pm 0.88)\times 10^{-3}$. 
This will allow us to make a parallel analysis 
as the $\chi_{c0}$ decays. By fitting the data, we obtain
\be
\label{para-value-2}
r=0.24\pm 0.29, \ \ \ R=1.09\pm 0.21, \ \ \  g_0=0.26\pm 0.06 \ \mbox{GeV}^{1/2}.
\ee
Again, the central values can be obtained by solving the corresponding equations, 
and large uncertainties for the fitting results 
highlight the effects from the experimental errors. 
It is interesting that the parameters for $\chi_{c0}$ and $\chi_{c2}$ 
are not dramatically different from each other. Instead, they are quite 
consistent. 
In particular, in contrast with $\chi_{c0}$ decays, 
the calculation shows that the OZI-rule is better respected in $\chi_{c2}$ decays. 
We also note that with a factor 3 reduction of the data errors in both cases, 
the uncertainty of $r$ can be improved to the second decimal place.

With the central values of the fitted parameters, we predict 
the branching ratios for $\omega\phi$ and $\rho\rho$ channels. 
Similar to $\chi_{c0}$, the channel $\omega\phi$ is relatively suppressed 
with $BR_{\chi_{c2}\to \omega\phi}=0.24\times 10^{-3}$. 
The branching ratio for $\rho\rho$ is found sizeable with a value of 
$BR_{\chi_{c2}\to \rho\rho}=2.41\times 10^{-3}$. We also present the estimated 
errors for these two channels in Table~\ref{tab-1}.

ii) $\chi_{c0,2}\to PP$

The same analysis can be applied to the $\chi_{c0,2}$ decays into pseudoscalar 
meson pairs. We will adopt the same notations as Part (i) for the parameters. 
But one should keep in mind that they do not necessarily have the same values. 
For $I=0$ channels, i.e. $\chi_{c0,2}\to \eta\eta$, 
$\eta^\prime\eta^\prime$, and $\eta\eta^\prime$, 
the mixing of $\eta$ and $\eta^\prime$ are defined as
\bea
\eta & = & \cos\alpha n\bar{n} -\sin\alpha s\bar{s} \nonumber\\ 
\eta^\prime & = & \sin\alpha n\bar{n} + \cos\alpha s\bar{s} \ ,
\eea
where $\alpha= 54.7^\circ + \theta_P$ and $\theta_P$ is the 
octet-singlet mixing angle in the SU(3) flavour basis.
We do not consider possible glueball components mixed within 
the $\eta$ and $\eta^\prime$ wavefunctions at this moment though 
they can be included. 
Quite directly, the transition amplitude of
Eq.~\ref{trans-1} can be reduced by requiring $x_i=x_j=0$ and $t=0$:
\bea
\label{pseudo}
\langle \eta\eta | \hat{V}| gg\rangle 
&=& g_0^2 [\cos^2\alpha + R^2\sin^2\alpha+ r(\sqrt{2}\cos\alpha -R\sin\alpha)^2]   \nonumber\\
\langle \eta^\prime\eta^\prime |\hat{V} | gg\rangle 
&=& g_0^2[\sin^2\alpha + R^2\cos^2\alpha + r(\sqrt{2}\sin\alpha +R\cos\alpha)^2]  \nonumber\\
\langle \eta\eta^\prime |\hat{V} | gg\rangle 
&=& g_0^2 r [(1-R^2/2)\sin 2\alpha +\sqrt{2} R \cos 2\alpha]  \nonumber\\
\langle K^+K^- |\hat{V} | gg\rangle 
&=& g_0^2 R \nonumber\\
\langle \pi^+\pi^- |\hat{V} | gg\rangle 
&=& g_0^2 \ ,
\eea
where the transition potential $\hat{V}$ is a compact form for 
$V_0+V_1+V_2+V_3$.

Experimental data for $\eta\eta$, $K^+K^-$ 
and $\pi\pi$ are available~\cite{bes-98i,bes-03c,pdg2004}, which will allow us to constrain 
the parameters and examine the model. 
We first determine $g_0^2$ in $\chi_{c0,2}\to \pi\pi$. Then, 
by taking the ratio between $K^+K^-$ and $\pi\pi$, 
we determine the SU(3) breaking parameter $R$:
\be
R=\left[\frac{3p_\pi {\cal F}^2(p_\pi) \Gamma_{K^+K^-}}
{2 p_K{\cal F}^2(p_K) \Gamma_{\pi\pi}}\right]^{1/2} \ ,
\ee
where factors 2 and 3 are the weighting factors for the charged kaon and pion 
decay channels. 
Substituting $\Gamma_{K^+K^-}=6.0\times 10^{-3}\Gamma_{tot}$ 
and $\Gamma_{\pi\pi}=7.4\times 10^{-3}\Gamma_{tot}$~\cite{bes-98i,pdg2004} 
into the above equation, we have $R=1.06$, which is in good agreement 
with the range found for $\chi_{c0}\to VV$. 

\begin{figure}
\begin{center}
\epsfig{file=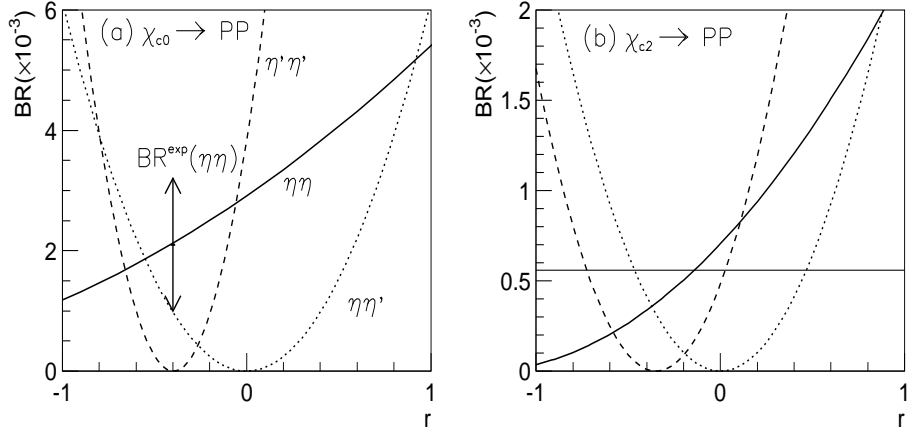, width=14cm,height=8.cm}
\caption{Branching ratios for the decays of $\chi_{c0}$ (Fig.(a)) 
and $\chi_{c2}$ (Fig.(b)) into 
$\eta\eta $ (solid curve), $\eta^\prime\eta^\prime $ (dashed curve), 
and $\eta\eta^\prime$ (dotted curve) in terms of the SU(3) parameter $r$. 
The two-direction arrow in (a) denotes the central value of 
$BR^{exp}_{\chi_{c0}\to\eta\eta}=(2.1\pm 1.1)\times 10^{-3}$, 
while the length marks the error bars. The thin straight line in (b) 
labels the upper limit 
$BR^{exp}_{\chi_{c2}\to\eta\eta} <1.5\times 10^{-3}$~\cite{bes-03c}.
}
\protect\label{fig-2}
\end{center}
\end{figure}

With the data for $\chi_{c0}\to \eta\eta$, we, in principle, 
can determine $r$ via the first equation in Eq.~\ref{pseudo}. 
However, we find that within the large uncertainties 
of the present data, 
$BR_{\chi_{c0}\to \eta\eta}=(2.1\pm 1.1)\times 10^{-3}$, 
the determination of parameter $r$ also possesses large uncertainties. 
To show this problem, we
plot the branching ratios for $\chi_{c0}\to \eta\eta$, $\eta^\prime\eta^\prime$ 
and $\eta\eta^\prime$ in terms of $r$ in Fig.~\ref{fig-2}(a). 
The solid line is for $BR_{\chi_{c0}\to \eta\eta}$, which exhibits 
a slow change with $r$. The arrow acrosses the central value of the 
experimental data, and marks the error bars. Therefore, the slow variation 
of $BR_{\chi_{c0}\to \eta\eta}$ makes it difficult to determine $r$. 
Interestingly, it shows that $BR_{\chi_{c0}\to \eta^\prime\eta^\prime}$ 
and $BR_{\chi_{c0}\to \eta\eta^\prime}$ are both fast changing functions 
of $r$. In other words, one can determine parameter $r$ by 
measuring these two branching ratios, and matching the pattern of the curves 
in a narrow region of $r$. 
Meanwhile, an improved measurement of $\eta\eta$ channel 
is also strongly recommended. 

For $\chi_{c2}\to PP$, there are also experimental data available 
for three channels: $BR_{\chi_{c2}\to K^+K^-}=(9.4\pm 2.1)\times 10^{-4}$, 
$BR_{\chi_{c2}\to \pi^+\pi^-}=(1.77\pm 0.27)\times 10^{-3}$, and 
$BR_{\chi_{c2}\to \eta\eta} < 1.5\times 10^{-3}$. 
However, notice that the branching ratio for $\eta\eta$ is only an upper limit 
and quite large uncertainties are with the data, 
the parameters cannot be well-constrained. 
Similar to case of $\chi_{c0}\to PP$, we can determine 
$R$ and $g_0$ with the data for $\pi\pi$ and $K\bar{K}$: 
$ R=0.70, \ \ \ g_0=0.30$. 
In comparison with the previous cases, it shows large SU(3) symmetry breakings. 
A possible reason could be due to the poor status of the data. 

Similar to $\chi_{c0}$ decays, we plot the change of branching ratios 
of $\eta\eta$, $\eta^\prime\eta^\prime$ 
in terms of a range of $r$. The results 
are presented in Fig.~\ref{fig-2}(b). Taking the upper limit of $\eta\eta$ 
branching ratio as a constraint, the OZI-rule violation  
parameter $r$ at least has a value of $-0.16$. Note that 
the relative branching ratio magnitudes among these three channels 
vary fast. Additional experimental information about $\eta\eta^\prime$ 
or $\eta^\prime\eta^\prime$ will provide better constraint on $r$.

iii) $\chi_{c0,2}\to SS$

For $\chi_{c0,2}\to SS$, where $S$ denotes scalars, 
a direct $gg\to GG$ coupling will also contribute 
to the $I=0$ decay amplitudes, and an additional 
parameter $t$ has to be included (see Eq.~\ref{trans-1}). 
Therefore, new features may arise from the $I=0$ decay channels.
In particular, it should be interesting to investigate 
the sensitivity of the decay branching ratios to the 
structure of the $I=0$ scalars, i.e. $f_0^i$. 
Specific patterns arising from these channels will be useful 
for providing evidence for the presence of glueball components 
in the scalar meson wavefunctions. 

In contrast, the decays of $\chi_{c0,2}\to a_0(1450) a_0(1450)$, 
and $K_0^*(1430)\bar{K_0}^*(1430)$ will be free of 
the interferences from the $gg\to GG$ transitions due to the isospin 
conservation. 

To proceed, we will focus on the decays of $\chi_{c0}\to f_0^if_0^j$. 
In principle, all the decay channels into any two of these scalars are 
allowed except for $f_0^1f_0^1$, of which the threshold 
is slightly above the $\chi_{c0}$ mass.
We will adopt the wavefunctions based on the glueball-$Q\bar{Q}$ mixings~\cite{close-amsler,close-kirk}, 
which are examined in the $J/\psi$ hadronic decays~\cite{close-zhao-f0}. 
As shown in Ref.~\cite{close-zhao-f0}, the branching ratios for 
$J/\psi\to V f_0^i$ exhibit specific patterns, which cannot be explained by 
simple $Q\bar{Q}$ mixings within the scalars. 
It is also found in Ref.~\cite{close-zhao-f0} 
that strong OZI-rule violation effects could contribute to the transition 
amplitudes. 

Due to lack of data, the 
parameters cannot be explicitly determined. 
However, based on our experiences from the first two parts, 
we can assume that the SU(3) flavour symmetry breaking is negligible. 
Thus, we fix $R=1$ in the calculations as an approximation. 
Also, for $gg\to GG$ coupling, we assume that it has the same strength 
as the SOZI transitions, i.e. $t=1$. We are then left with parameters $g_0$ 
and $r$ for which, in principle, experimental constraints are needed. 
But we can still proceed one step further to consider 
the branching ratio fractions, where parameter $g_0$ cancels.

\begin{figure}
\begin{center}
\epsfig{file=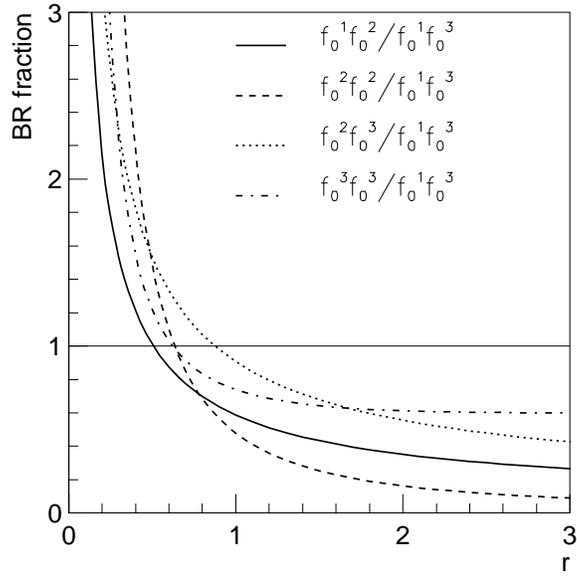, width=10cm,height=10.cm}
\caption{Branching ratio fractions for $\chi_{c0}$ decays 
into $f_0$ pairs in terms of the OZI-rule violation parameter $r$. 
The thin straight line denotes ratio unity and separates the two regions for  
the ratios greater/smaller than one. 
}
\protect\label{fig-3}
\end{center}
\end{figure}

Taking $\chi_{c0}\to f_0^1f_0^3$ as reference, we plot in Fig.~\ref{fig-3} 
the branching ratio fractions of $BR_{f_0^if_0^j}/BR_{f_0^1f_0^3}$ 
in terms of $r$ in a range of $0<r<3$. 
Interestingly, it shows that the branching ratio fractions 
are separated into two distinguished regions. 
In the limit of $r\to 0$, the branching ratio of 
$\chi_{c0}\to f_0^1f_0^3$ becomes the smallest one, 
and $\chi_{c0}\to f_0^2f_0^2$ is the largest. 
However, at the region of $r>1$, the branching ratio of 
$\chi_{c0}\to f_0^1f_0^3$ becomes the largest. 

In Refs.~\cite{close-zhao-f0,zhao-zou}, large OZI-rule violations 
are found in association with the scalar meson production in $J/\psi$ decays. 
Therefore, we conjecture that large contributions from 
the DOZI processes will be present in $\chi_{c0}\to f_0^if_0^j$. 
Based on this, the branching ratio fractions at $r>1$ 
strongly suggest 
that $\chi_{c0}\to f_0^1f_0^3$ will be the largest decay channel. 
Notice that $f_0^1$ couples strongly to $K\bar{K}$, while 
$f_0^3$ strongly to $\pi\pi$, one might be able to see rather clear 
evidence for these two states in e.g. 
$\chi_{c0}\to f_0^1f_0^3\to K^+K^-\pi^+\pi^-$ at BES~\cite{bes-chi-c0}. 

Due to lack of data for $\chi_{c2}\to SS$, we do not discuss 
this channel in this work.

To summarize,
by distinguishing the SOZI and DOZI processes 
and including a direct glueball production mechanism    
in the transition amplitudes, the overall available data 
can be accounted for. Since these basic transition processes 
have been factorized out, we can study their correlations 
among all the decay channels. 
We find that the OZI-rule violations play quite different
roles in the production of $VV$, $PP$, and $SS$ in the $\chi_{c0,2}$ decays. 
This may explain that pQCD approaches 
systematically underestimate 
the data~\cite{anselmino,chernyak,cao-huang-wu}\footnote{We notice that in hep-ph/0506293, an improved 
result is obtained by Luchinsky at leading twist}. 
On the other hand, a recent study based on the $^3P_0$ quark pair 
creation model shows that the nonperturbative mechanism 
still plays important roles in $\chi_{c0}\to \phi\phi$~\cite{zhou-ping-zou}.
In this approach, a coherent description of both pQCD favored and unfavored 
mechanisms allows us not only to isolate their contributions, but also 
to investigate their correlated interferences. 
Such correlations produce specific patterns for 
the branching ratio magnitudes of different 
channels, and can be highlighted by experimental data.

For $\chi_{c0,2}\to VV$, we find that the branching ratio of 
$\chi_{c0}\to \omega\phi$ 
is crucial for determine the role of the OZI-rule violations, 
and for $\chi_{c0,2}\to PP$, an improved measurement 
of $BR_{\chi_{c0}\to \eta\eta}$ will clarify the correlations 
from the OZI violations. For $\chi_{c0}\to f_0^i f_0^j$, 
we find that large OZI-rule violations will lead to suppressions on  most of 
the branching ratios except $f_0^1f_0^3=f_0(1710)f_0(1370)$. 
With an analogue to the large OZI-rule violations in the $f_0$ scalar productions 
in $J/\psi\to V f_0$, we would expect that large OZI-rule violations 
would occur in $\chi_{c0}\to f_0^i f_0^j$. Therefore, 
an observation of sizeable branching ratio for 
$f_0^1f_0^3=f_0(1710)f_0(1370)$ will be a strong signal for 
the glueball-$Q\bar{Q}$ mixings within the scalars. 
In contrast, small OZI violations will lead to relatively smaller
branching ratios for $f_0^1f_0^3$. 
If this occurs, we will then have difficulty to understand 
the large OZI violations in $J/\psi\to Vf_0^i$~\cite{close-zhao-f0,zhao-zou}. 
Experimental data from BES Collaboration can provide an important 
test of this approach, and also put a more stringent constraint 
on the model parameters. We expect that 
a confirmation of this will also provide an 
additional evidence for the glueball-$Q\bar{Q}$ mixings within those 
scalars.

The author thanks F.E Close and B.S. Zou for useful comments on this work, 
and Z.J. Guo and C.Z. Yuan for useful communications 
about the BES experiment. 
This work is supported,
in part, by grants from
the U.K. Engineering and Physical
Sciences Research Council Advanced Fellowship (Grant No. GR/S99433/01),
and
the Institute of High Energy, Chinese Academy of Sciences.


\begin{table}[ht]
\begin{tabular}{c|c|c}
\hline
Decay channels 
& $BR_{\chi_{c0}\to VV}(\times 10^{-3})$ 
& $BR_{\chi_{c2}\to VV}(\times 10^{-3})$  \\[1ex]
\hline
$\phi\phi$ & $1.0\pm 0.6$ & $2.00\pm 0.82$ \\[1ex]
$\omega\omega$ &  $2.29\pm 0.71$ & $1.77\pm 0.59$  \\[1ex]
$K^*\bar{K^*}$ & $1.78\pm 0.48$ & $4.86\pm 1.04$  \\[1ex]
$\omega\phi$ & {\bf 0.45 \ \ (1.07)} & {\bf 0.24 \ \ (0.65)}  \\[1ex]
$\rho\rho$ & {\bf 1.88 \ \ (1.80)} & {\bf 2.41 \ \ (2.22)}  \\[1ex]
\hline
\end{tabular}
\caption{ The branching ratios for $\chi_{c0,2}\to VV$. 
The first three rows ($\phi\phi$, $\omega\omega$ and $K^*\bar{K}^*$) are 
experimental data from BES~\cite{bes-99b,bes-04,bes-05}, 
while the last two rows, highlighted by bold faces, 
are the model predictions for $\omega\phi$ and $\rho\rho$ channels. 
The numbers in the brackets are the root mean square errors 
estimated with Eqs.~\ref{para-value} and \ref{para-value-2}, respectively. }
\label{tab-1}
\end{table}


\end{document}